\documentclass[]{iopart}

\usepackage[dvips]{graphicx}
\usepackage{setstack}
\usepackage{amsfonts}

\begin{document}

\title{The Interactive Minority Game: Instructions for Experts}
\date{15 August 2002}

\author{Peter Ruch, Joseph Wakeling\footnote[1]{{\tt jwakeling@webdrake.net}, {\tt joseph.wakeling@unifr.ch}} and Yi-Cheng Zhang\footnote{{\tt yi-cheng.zhang@unifr.ch}}}
\address{Institut de Physique Th\'eorique, Universit\'e de Fribourg P\'erolles, CH-1700 Fribourg, Switzerland}
\address{{\tt http://www.unifr.ch/econophysics/minority/game/}}
\author{15 August 2002}

\begin{abstract}
The \emph{Interactive} Minority Game (IMG) is an online version of the traditional Minority Game in which human players can enter into competition with the traditional computer-controlled agents.  Through the rich (and, importantly, analytically understood) behaviour of the MG, we can explore humans' behaviour in different kinds of market---crowded, efficient, critical---with a high degree of control.  To make the game easily understandable even to those who are encountering it for the first time, we have presented the game with a rather simplified interface; in this working paper we explain the underlying technical aspects for those who have experience with the traditional MG.
\end{abstract}

\section{Introduction}

The Minority Game (MG), introduced by Challet and Zhang~\cite{CZ97} as a generalisation of Arthur's `El Farol' problem~\cite{A94}, has gained much popularity as a simple yet realistic model of the workings of financial markets.  A great variety of work has since been done on the MG~\cite{minority}, leading eventually to an analytic solution~\cite{MCZ00}; these insights have proven particularly useful in justifying the MG as a realistic model of real market situations~\cite{CMZ00}.

The Interactive Minority Game (IMG) proposes to further this use of the MG in modelling real markets by using the MG to try to gain insight into the behaviour of humans in a simplified market situation.  Individual humans compete against the traditional computer-controlled agents, whose characteristics can be fine-tuned through our understanding of the traditional MG: the human player can be presented with a very inefficient and easy-to-predict market, or with one that is much more complex and difficult.

In creating an interface for the game, our emphasis was on simplicity and ease of use: the game should be playable by anyone, from a seasoned MG researcher to someone who comes across the website at random.  A number of aspects of the interface may therefore seem a little strange to those who have experience of the MG, and so in this paper we explain the mechanical workings-behind-the-scenes that go into the IMG: firstly, the interface, and secondly, the setup and parameters behind the different choices of game.

\section{An interactive interface}

Fundamentally, the Interactive Minority Game is a minority game like any other.  A group of $N$ agents must choose between two actions, $+1$ (`buy') or $-1$ (`sell'), with the intention to make the opposite choice to the majority of agents; agents' decisions are based on the knowledge of which was the minority or majority action of the last $M$ turns.  However, in order to make the game a little easier to play, we have had to change the presentation (though not the underlying game!) a little for the sake of the human player.  The human player not only has the advantage of a greater memory (denoted by $M_{\ast}$) than the computer players, but is also given a choice between two different methods of presentation for the information about the game history; in addition, rather than asking the human player for an action (`buy' or `sell'), we ask them for a prediction on the direction the market is going to take, which corresponds to an action in the underlying game.  In this section we explain (and, we hope, justify) in detail all these aspects of the interface.

\subsection{\label{sec:2-1}Views of the market: binary and price-chart}

Imagine being confronted by the string of 1's and 0's that is given to the regular agents!  It is not a particularly pleasant thought, and the human player will clearly demand more.  We have provided two different solutions to the problem.

Before we continue, let us recall a few aspects of the MG.  At each time step $t$, each agent $i$ takes an an action $a_{i}(t)$, which allows us to define the $global$ action $A(t)$:

\begin{equation}
A(t) = \sum_{i}a_{i}(t)
\end{equation}

\begin{figure}[t]
\includegraphics[width=\textwidth]{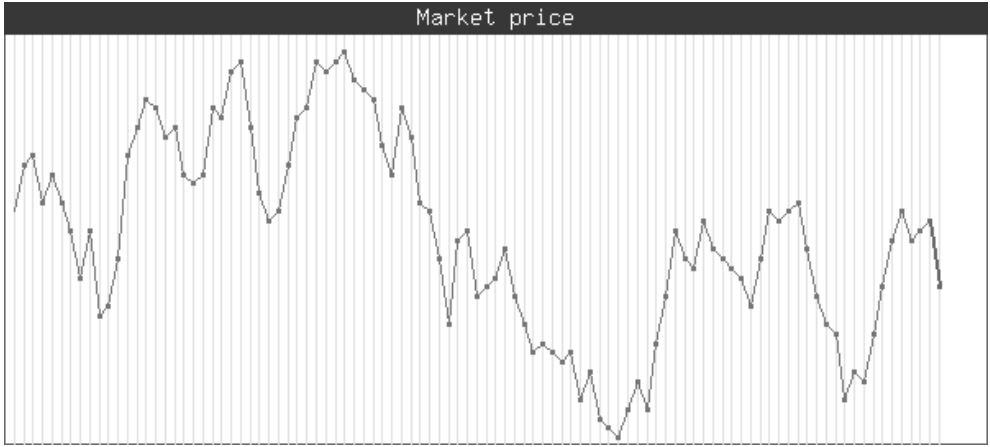}
\caption{\label{fig:2}~The price-chart view of the game.  The price changes proportionally to the global action $A(t)$ of agents (Eq.~\ref{eq:2}); the player must predict whether the market will rise or fall.  Note that the last action is highlighted for the player.}
\end{figure}

This allows us to define the first, simpler view of the game history: the \emph{binary viewpoint}, a set of spins corresponding to $\mathsf{sign}\left( A(t)\right)$ in the the last $M_{\ast}$ turns of the game.  This is of course identical in information content to the aforementioned string of 1's and 0's, but looks nicer!  The more interesting viewpoint follows from an idea suggested by Johnson \emph{et.\ al.}~\cite{JHHZ00} of a price level, $P(t)$, in the MG, defined by

\begin{equation}
\label{eq:2}
P(t+1) = P(t) + A(t)/\lambda
\end{equation}

where $\lambda$ is the market liquidity.  We have used this in our \emph{chart viewpoint}, literally a price-chart recording the past values of $P(t)$ (see Fig.~\ref{fig:2}).

Clearly, the binary viewpoint is more suited to games where the player's gain is also binary, i.e.\ where $g_{i}(t) = -a_{i}(t)\ \mathsf{sign}\left( A(t)\right)$, and the price-chart view is more suited to games with a linear payoff, $g_{i}(t) = -a_{i}(t)A(t)$.  This is the practise we have followed in our ready-made games.  However, the option to create a custom game allows the user to ignore our advice if s/he feels like it!

\subsection{\label{sec:predict}Predictions, bets and actions}

The reader may have noticed that, rather than being asked to make a choice (`buy' or `sell', $+1$ or $-1$, \ldots), the human player is asked to `\emph{predict which way you think the market is going to go}'.  That is, the user is asked to \emph{predict the majority action}, and their own action (the opposite) is inferred from this---we ask the human player for a prediction $q_{\ast}(t) = \pm 1$ and an amount bet on this prediction, $\omega_{\ast}(t)\in\{1,3,5,7\}$ (their \emph{weight} or \emph{market impact}), and their action is then taken to be $a_{\ast}(t) = -\omega_{\ast}(t)q_{\ast}(t)$.

We apologise if this is perhaps a little confusing for those familiar with the MG!  Our concern, as we noted in the Introduction, was to make the game simpler for those who have never encountered the MG before.  `What way do you think the market will go?' seemed to us a simpler question than, say, `Would you like to buy or sell (do the opposite of what you think everyone else will do)?'  In addition, those unused to the MG might reasonably ask, like Andersen and Sornette in a recent paper~\cite{AS02}, questions such as, `Why does buying hurt me if the market keeps rising?'  We would rather avoid such problems!

In summary: although it is a little hidden from view, the player still has to try and perform the minority action. {\tt ;-)}



\subsection{Nuts and bolts of the implementation}

The game interface consists of a series of PHP files that interact with an underlying C program; the latter calculates the actions of all the computer-controlled agents, and calculates all the relevant and interesting quantities (such as predictability).  We hope that the reader will forgive us if we are coy about revealing any more (in particular, about revealing source code); we are willing to reveal the workings-behind-the-scenes but there are a few rabbits-up-our-sleeve that we would like to keep from players for now, so as to make the game more interesting!

\section{Mechanics and game parameters}

The Interactive Minority Game offers a number of possibilities for the human player to pick from.  As well as a variety of `standard', ready-made games, there is the option to create a custom game by choosing from various parameters.  For those well-versed in MG lore, this could no doubt do with some extra explanation.  Parameters particular to the human player are denoted by a subscript asterisk ($\ast$).

\subsection{Game parameters and their effects}

To begin with, let us summarise the basic parameters of the MG: we have a homogeneous population of $N$ agents with memory $M$, each with $S=2$ strategies.  This gives us $P=2^{M}$ as the size of the space of histories given to agents, and we can then define the control parameter $\alpha = P/N$.  It is this parameter $\alpha$ which determines the type of market: a low value creates an extremely inefficient market, easily predictable by the smart human, while a high value of $\alpha$ creates a market whose behaviour is close to randomness.  The most efficient market (and therefore most difficult to play), i.e.\ the market with the minimal value of standard deviation $\sigma^{2}/N$, lies somewhere in the middle at the critical point $\alpha_{\mathrm{c}} = 0.3374\dots$~\cite{SMR99,CM99,C00}.  For the purposes of our games (custom or standard), we have chosen to make $M$ and $\alpha$ variables, with other parameters being calculated from these two.

As we have already noted, the human player enjoys the benefit of a larger memory, $M_{\ast} > M$, than the computer-controlled agents; we have chosen to set $M_{\ast} = 25$ for the binary view and $M_{\ast} = 50$ for the price-chart view.  We have also added a parameter particular to the IMG for the human player: the player's \emph{weight} or \emph{market impact}.  Recall from \S\ref{sec:predict} that at each time $t$ the player is asked for a prediction $q_{\ast}(t)$ on the direction the market will take, and the amount they bet on this, $\omega_{\ast}(t)\in\{ 1, 3, 5, 7\}$.  Their action is then taken to be,
\begin{equation}
a_{\ast}(t) = -\omega_{\ast}(t)q_{\ast}(t)
\end{equation}

Put simply, this amounts to the human player taking an action $a_{\ast}(t) = \pm\omega_{\ast}(t)$ rather than $\pm 1$; i.e., the player chooses to have a market impact equivalent to that of $\omega_{\ast}(t)$ normal agents.  The player's gain per turn is then also proportional to $\omega_{\ast}(t)$.  Depending on the market phase, this can be an advantage or a disadvantage: it brings the potential of a greater reward if the player can correctly predict the market, but with the potential of a far greater loss if s/he cannot.

We have also used the concept of \emph{minimum weight}, $\omega_{\ast}^{\mathrm{min}}$, where we require that $\omega_{\ast}(t) \geq \omega_{\ast}^{\mathrm{min}}$.  We can force the human player to act on a larger scale than the typical agent---equivalent, perhaps, to the actions of a large investment bank rather than an individual trader.

The other new parameter introduced, for custom games, is the length of the \emph{introductory period}, $T_{\mathsf{int}}$.  This is the length of time that the computer-controlled agents play by themselves before the human player joins the game.  A long value allows the agents to adapt to the statistically stationary state of the game, while if $T_{\mathsf{int}}$ is low, the players start off ``stupid'', not knowing which of their strategies is better.  In the more difficult games (efficient markets), a such a low value is essential as a high value of $T_{\mathsf{int}}$ makes it completely impossible for a human to compete!

\subsection{Parameter choices}

The IMG allows the player both to pick from a variety of `standard', ready-made games and to create their own custom game.  The total choices available to the player are as follows: their \emph{minimum weight}, $\omega_{\ast}^{\mathrm{min}}$; the \emph{market phase}, determined by $\alpha$; the \emph{agent memory}, $M$; the display mode (see \S \ref{sec:2-1}); the payoff (binary or linear); and, in the custom game, the length of the introductory period, $T_{\mathsf{int}}$.

The {\bf custom game} allows the choice of a full range of parameters, including some that are not used in our standard games.  Minimum weight can be Small ($\omega_{\ast}^{\mathrm{min}} = 1$), Medium ($\omega_{\ast}^{\mathrm{min}} = 3$) or Large ($\omega_{\ast}^{\mathrm{min}} = 5$) absolute values; the market phase varies from extremely crowded ($\alpha = 0.05$) to crowded ($0.1$ or $0.2$), the critical point ($0.34$) and anticrowded ($1.0$ or $2.0$).  $M$ can be 1, 2, 3, 4, 6 or 8; the introductory period can be the full period ($T_{\mathsf{int}} = 1000$), or it can be short (200) or very short (5).

The {\bf standard games} were set up to create a variety of different market experiences for the human player.  \emph{Beginners} is exactly that: a nice and easy game, with $M=2$, $\alpha = 0.05$ and $\omega_{\ast}^{\mathrm{min}} = 1$.  This is designed simply as a friendly introduction to the IMG and should not be taken too seriously!

The \emph{Newbie Traders} game is a little trickier, with with $\omega_{\ast}^{\mathrm{min}} = 1$ and $M = 2$ as before, but now with $\alpha = 0.2$.

\emph{Day Traders} is again trickier, with $M=3$,  $\alpha = 0.05$ and $\omega_{\ast}^{\mathrm{min}} = 1$, while \emph{Chartists} ups the ante still further with $\alpha = 0.2$.

\emph{Wall Street Gurus} throws the player into a difficult market at the critical point, with $M=4$, $\alpha = 0.34$ and $\omega_{\ast}^{\mathrm{min}} = 1$.

\emph{Mission Impossible} is literally that---the player must struggle with a large $\omega_{\ast}^{\mathrm{min}} = 3$ and an anticrowded market ($\alpha = 2.0$) full of very intelligent agents ($M=6$).

All standard games use the price-chart view with a linear payoff.

\section{What are we looking for in the IMG?}

While clearly the fundamental aim of the IMG (apart from being a fun distraction for the econophysics community!)\ is to investigate the behaviour of humans in market situations, we should perhaps specify a bit more precisely what we are looking for.  Here are a few of the questions we have in mind:

\begin{itemize}
\item What sort of performance do we see from humans in different types of market (crowded, anticrowded, critical\ldots)?
\item How do humans make their decisions?  Do they stick to one strategy or vary their behaviour often?
\item How much information do human players give to the market?
\item Do they take into account their market impact?
\end{itemize}

In conclusion, we hope that the IMG, while a very simple and easy game, might provide a fascinating window into what is actually going on inside the heads of real, human, economic agents---surely one of the ultimate econophysical goals!

\end{document}